\documentclass[aps,prl,preprintnumbers,amsmath,amssymb,latexsym,array,enumerate,letter,twocolumn,superscriptaddress]{revtex4}

\usepackage{amssymb}
\usepackage{amsmath}
\usepackage{epsfig}
\usepackage{hyperref}
\usepackage{breakurl}
\usepackage{xcolor}

\makeatletter
\def\simgt{\mathrel{\lower2.5pt\vbox{\lineskip=0pt\baselineskip=0pt
           \hbox{$>$}\hbox{$\sim$}}}}
\def\simlt{\mathrel{\lower2.5pt\vbox{\lineskip=0pt\baselineskip=0pt
           \hbox{$<$}\hbox{$\sim$}}}}
\makeatother

\newcommand{\be}{\begin{equation}}
\newcommand{\ee}{\end{equation}}
\newcommand{\bea}{\begin{eqnarray}}
\newcommand{\eea}{\end{eqnarray}}
\newcommand{\beq}{\begin{eqnarray}}
\newcommand{\eeq}{\end{eqnarray}}

\def\lsim{\mathrel{\rlap{\lower4pt\hbox{\hskip1pt$\sim$}}
     \raise1pt\hbox{$<$}}}         
\def\gsim{\mathrel{\rlap{\lower4pt\hbox{\hskip1pt$\sim$}}
     \raise1pt\hbox{$>$}}}         

\begin{document}

\widetext

\title{Using LISA-like Gravitational Wave Detectors to Search for Primordial Black Holes}


\author{Huai-Ke Guo}
\affiliation{
CAS Key Laboratory of Theoretical Physics, Institute of Theoretical Physics, \\
Chinese Academy of Sciences, Beijing 100190, China}
\author{Jing Shu}
\affiliation{
CAS Key Laboratory of Theoretical Physics, Institute of Theoretical Physics, \\
Chinese Academy of Sciences, Beijing 100190, China}
\affiliation{
School of Physical Sciences, University of Chinese Academy of Sciences, Beijing 100190, P. R. China
}
\affiliation{
CAS Center for Excellence in Particle Physics, Beijing 100049, China
}

\author{Yue Zhao}
\affiliation{Tsung-Dao Lee Institute, and Department of Physics and
Astronomy, Shanghai Jiao Tong University, Shanghai 200240}
\affiliation{ Michigan Center for Theoretical Physics, University of
Michigan, Ann Arbor, MI 48109 }

\begin{abstract}
Primordial black holes (PBH), which can be naturally produced in the
early universe, remain a promising dark matter candidate . They can
merge with a supermassive black hole (SMBH) in the center of a
galaxy and generate a gravitational wave (GW) signal in the favored
frequency region of LISA-like experiments. In this work, we initiate
the study of the event rate calculation for such extreme mass ratio
inspirals (EMRI). Including the sensitivities of various proposed GW
detectors, we find that such experiments offer a novel and
outstanding tool to test the scenario where PBHs constitute
(fraction of) dark matter. The PBH energy density fraction of DM
($f_\text{PBH}$) could potentially be explored for values as small
as $10^{-3} \sim 10^{-4}$. Further, LISA has the capability to
search for PBH masses up to $10^{-2} \sim 10^{-1} M_\odot$. Other
proposed GW experiments can probe lower PBH mass regimes.

\end{abstract}


\maketitle \noindent{\bfseries Introduction.} Dark matter (DM)
comprises about 27$\%$ of the energy density in our current
universe~\cite{Ade:2015xua}. However the identity of DM remains a
mystery. It may be particles beyond the Standard Model, where
popular choices are Weakly Interacting Massive Particle and axion.
Primordial black holes (PBH) are also a promising candidate with a
wide allowed mass range ~(for a PBH review, please see
e.g.~\cite{Carr:2009jm}). There have been a lot of efforts to study
the fraction of DM as PBH, e.g. using gravitational
lensing~\cite{Nemiroff:2001bp,1994ApJ424550D,Wilkinson:2001vv,Niikura:2017zjd,Tisserand:2006zx,Wyrzykowski:2011tr,Allsman:2000kg,Alcock:1998fx,Griest:2013aaa},
the CMB temperature anisotropies and polarizations
~\cite{Ricotti:2007au,Chen:2016pud}, etc. The validity as well as
astrophysical uncertainties of these constraints are still under
debate, ~\cite{Ali-Haimoud:2016mbv,Green:2017qoa} and thus it is
interesting to explore this possibility through new and independent
measurements.


The detection of the gravitational wave (GW) events from black hole
binaries by the LIGO and Virgo collaborations
~\cite{Abbott:2016blz,Abbott:2016nmj,Abbott:2017vtc} has begun the
era of GW astronomy. GW observations provide a novel method to study
the universe. Many GW detectors have been proposed (see
Ref~\cite{Moore:2014lga} for a review). In particular, Laser
Interferometer Space Antenna (LISA), which aims for a much lower
frequency regime than that of LIGO-like ground-based detectors, has
been approved recently ~\cite{Audley:2017drz}. One major scientific
goal of LISA is to measure the GW produced by the merger of a SMBH
and a compact object (CO), such as a neutron star, white dwarf or
stellar BH. In such EMRIs ~\cite{Amaro-Seoane:2014ela}, GW
frequencies typically range from $10^{-4}$ to $ 1\ \text{Hz}$ for
SMBH masses between $10^4 M_{\odot}$ and $10^7 M_{\odot}$. Once such
events are observed, the intrinsic parameters of the binary system
can be measured in high precision~\cite{Barack:2003fp} due to the
long-lasting inspiral process before merging.

Aside from their significant impacts for astronomy, the observation
of GWs may also open a new avenue to study the possibility of PBHs
playing the role of DM. Especially,
Ref.~\cite{Bird:2016dcv,Sasaki:2016jop,Carr:2016drx,Eroshenko:2016hmn,Clesse:2016ajp,Orlofsky:2016vbd,Nishikawa:2017chy}
study the interesting question of whether the BHs detected by LIGO
can be PBHs which form a non-trivial fraction of DM. Using LIGO and
LISA to probe extremely small mass PBH is studied
in~\cite{Takhistov:2017bpt}. For PBHs with mass of $\mathcal{O}(10)\
M_\odot$, it is hard to distinguish them from stellar BHs. However,
LIGO is not ideal to probe other PBH mass ranges, either due to the
shifted frequency region or reduced magnitude of GW radiation. On
the other hand, the mergers between PBHs and SMBHs produce GWs in
the favored frequency regions of LISA-like experiments. Such
frequencies are mainly determined by SMBH mass and are independent
of PBH mass. This indicates that, unlike LIGO, we potentially have
the access to a vast mass range of PBHs, which lies outside the mass
window of astrophysical COs. Therefore, observation of these events
may be used to claim the discovery of PBHs. Moreover, the DM profile
peaks at the center of a galaxy, indicating the possibility of a
large number density of PBHs in the neighborhood of a SMBH. This may
induce a significant EMRI rate caused by PBH-SMBH mergers.

In this letter, we carry out the first study of the event rate
estimation for PBH-SMBH mergers, taking into considerations the
sensitivities of different experiments. In the next section, we
outline the essential ingredients for the calculation. Then we
calculate each of them in the later sections. After that, we put
everything together and interpret the observable event rate for
different experiments as their capabilities to probe PBH-as-DM
scenarios. We find these experiments provide us a powerful tool to
study a large unexplored parameter space. Not only could the
sensitivity to $f_\text{PBH}$ be as good as $10^{-3} \sim 10^{-4}$,
but also the lower limit of PBH masses that can be probed is
potentially far from the astrophysical CO mass region. This could be
used to discover PBH from these GW experiments.


\vspace{0.2cm}

\noindent{\bfseries Ingredients for EMRI Rate Calculation.} EMRI has
been carefully studied in the context of astrophysics. In
particular, the merger rate between SMBH and astrophysical COs has
been calculated. Let us first summarize the key ingredients in this
calculation.

The event rate observed by a GW detector can be written as,
\begin{eqnarray}\label{eq:master}
\Gamma = \int  \mathcal{R}(M, \mu)  \left(\frac{dn(M,z)}{d M}
   d M \right) \left(p(s,z) ds \right) \left( \frac{d V_c}{dz} dz \right),
\label{eq:rateall}
\end{eqnarray}
where $\mathcal{R}(M,\mu)$ is the intrinsic EMRI rate in a galaxy
hosting a SMBH with mass $M$. The mass of the CO is $\mu$. The
${dn(M, z)}/{d M}$ and $p(s,z)$ are the mass spectrum and spin, $s$,
distribution of SMBHs. They are functions of redshift $z$ due to the
evolution of galaxies. If one only focuses on late times,
$z$-dependence may be approximately removed. From the popIII
model~\cite{Klein:2015hvg}, most of the SMBHs within the LISA range,
i.e. with mass comparable or smaller than $10^7\ M_\odot$, are
expected to have near maximal spins ~\cite{Babak:2017tow}. Further,
EMRI rates are calculated with various spin distributions, and the
difference appears to be less than 10$\%$. Thus in the following
discussion, we fix $s=0.999$.

In addition $\left( \frac{d V_c}{dz}dz\right)$ is the comoving
volume integral as a function of $z$. Since the GW strength
decreases when distance increases, not all EMRI events are
detectable by a GW detector. Thus the sensitivity of an experiment
imposes a maximum $z$, $z_{max}$, as a function of ($M$, $s$,
$\mu$), the details of which we will discuss in later sections.

Among these ingredients, the most non-trivial is
$\mathcal{R}(M,\mu)$. The intrinsic EMRI rate can be calculated by
solving the Fokker-Planck equation, which describes the diffusion of
the CO distribution functions. The result is a function of the mass
and density of the CO. Although the precise result has yet to be
obtained by numerical calculations, qualitative estimation is
possible and agrees well with numerics~\cite{Hopman:2005vr}.

As far as is known, the detailed numerical calculation on
$\mathcal{R}$ is only done assuming COs are white dwarfs, neutron
stars and stellar BHs. It is important to derive a reasonable
estimation on intrinsic EMRI rate for PBHs whose mass and number
density are dramatically different from those of astrophysical COs.
We will follow the analysis in ~\cite{Hopman:2005vr} and present an
analytical formula to scale $\mathcal{R}$ for stellar BHs as a
function of the PBH's properties.

In the next few sections, we prepare the ingredients for the
calculation of Eq.(\ref{eq:rateall}). We first discuss the DM
profile, which determines the number density of PBHs near a SMBH.
Astrophysical empirical equations are applied to relate DM profiles
to SMBH masses. Then we review the calculation of the GW strain from
EMRIs. We show sensitivities of various GW detectors and discuss the
calculation of signal-to-noise-ratio (SNR). We also consider the
subtlety of how detector operation time affects the SNR estimation.
After that, we present a detailed analysis of how the intrinsic EMRI
rate scales as a function of PBH number density and mass. Last, we
put everything together to study the event rate for various GW
detectors.

\vspace{0.2cm}

\noindent{\bfseries Dark Matter Halo Profile.} The PBH-SMBH merger
rate highly depends on the number density of PBHs around SMBH. EMRIs
are mainly produced by COs within the radius of influence of the
SMBH ~\cite{1972ApJ...178..371P},
\begin{equation}\label{eq:RoI}
  r_h = \frac{G M}{\sigma^2}  = 2 \text{pc} \left( \frac{M}{3 \times 10^6 M_{\odot}}
  \right)^{1/2} ,
\end{equation}
where $\sigma$ is the velocity dispersion in the bulge, and the
following $M-\sigma$ relation
~\cite{Ferrarese:2000se,Gebhardt:2000fk,Tremaine:2002js} is applied:
\begin{equation}
M = 10^8 M_{\odot} \left(\frac{\sigma}{200 \text{km/s}}\right)^4 .
\end{equation}

Since $r_h$ is $\mathcal {O}(\textrm{pc})$, the EMRI rate is
sensitive to the DM energy density in the innermost region. While
collisionless N-body simulations of cold DM indicate a cuspy
profile~\cite{Dubinski:1991bm,Navarro:1995iw,Navarro:1996gj,Moore:1999gc},
a cored profile may be obtained if other effects, such as baryonic
feedback, are taken into consideration~\cite{Tulin:2017ara}. On the
other hand, assuming adiabatic growth of SMBHs, a spike around the
galactic center can be induced
~\cite{Gondolo:1999ef,Sadeghian:2013laa} and is more pronounced for
a Kerr SMBH~\cite{Ferrer:2017xwm}. Especially,
in~\cite{Nishikawa:2017chy}, a spike connected to the NFW profile is
used to study the PBH-PBH merger rate, which is enhanced as
expected. In this letter, we only use the NFW
profile~\cite{Navarro:1995iw,Navarro:1996gj} as an illustration and
note that cored (spiky) profiles may lead to smaller (larger) rates.

The NFW profile can be parametrized as
\begin{equation}
\rho(r) = \frac{\rho_s}{\frac{r}{R_s}(1 + \frac{r}{R_s})^2},
\end{equation}
where $\rho_s$ and $R_s$ are the characteristic density and scale
radius, respectively. The enclosed mass within a radius $R$
(equivalently, the dimensionless radius $c\equiv R/R_s$) is
\begin{eqnarray}
m_{\text{Halo}}
= \int_0^{R_{\text{max}}} 4 \pi r^2 \rho(r) d r
= 4\pi \rho_s R_s^3 g(c_{\text{max}}) ,
\end{eqnarray}
where the function $g(x)=\ln(1+x)-x/(1+x)$ is defined for later
convenience. Since $m_{\text{Halo}}$ diverges, a cutoff radius is
conventionally defined such that the enclosed average DM energy
density is 200 times the critical density of the universe $\rho_c$.
The DM halo profile can then be specified by the two parameters
$c_{200}$ and $M_{200}$, where $M_{200}$ is the enclosed DM halo
mass, and $c_{200}$ is the corresponding radius in units of $R_s$:
\begin{eqnarray}
\rho_s = \frac{200}{3} \frac{c_{200}^3}{g(c_{200})} \rho_c;
\quad
R_s =
\left[\frac{M_{200} }{4 \pi \rho_s g(c_{200})}\right]^{1/3} .
\end{eqnarray}
Further, at late times in the universe, i.e. at small redshift,
$c_{200}$ and $M_{200}$ can be related through the
concentration-mass relation~\cite{Dutton:2014xda},
\begin{eqnarray}\label{eq:DMProfile}
c_{200} = 10^{0.905} \left(\frac{M_{200}}{10^{12} h^{-1}
M_{\odot}}\right)^{-0.101} . \label{eq:c-m}
\end{eqnarray}
Here $h=0.673$ is the Hubble parameter at present time. The DM halo
can then be specified by a single parameter, chosen here as
$M_{200}$. Since Eq.~(\ref{eq:c-m}) only holds at small $z$, we
truncate the spatial integral in the rate calculation at a maximal
distance. More explicitly, we take $z \le 1$ ($r_0 \le 3.5
\text{Gpc}$).

Last, we need the connection between the halo mass $M_{200}$ and the
SMBH mass $M$. This is given in~\cite{Ferrarese:2002ct},
\begin{eqnarray}
\frac{M}{3 \times 10^6 M_{\odot}} \approx 3.3
\left(\frac{M_{200}}{10^{12} M_{\odot}}\right)^{1.65} .
\end{eqnarray}
Therefore, the DM halo profile can be expressed as a simple function
of the SMBH mass. We note that the total DM mass within $r_h$,
according to the above NFW profile, is $\sim 10^{-2}$ of the SMBH
mass. Thus the existence of DM can be treated as small perturbation.

\vspace{0.2cm}

\noindent{\bfseries Gravitational Wave Strain and SNR.} Modeling GW
emission from an EMRI system is non-trivial. Several formalisms have
been studied. For example, the numerical-kludge
model~\cite{Gair:2005ih,Babak:2006uv} is more accurate but
computationally expensive. The analytic kludge model
(AK)~\cite{Peters:1963ux,Barack:2003fp}, on the other hand, is
cheaper but at the price of accuracy. Within AK formalism, the two
ways to truncate the calculation are labeled as AKK and AKS, which
tend to give optimistic/conservative estimates of SNR. These two
choices characterize the uncertainties of the calculation. Last,
gravitational wave emission can also be approximately calculated for
circular and equatorial EMRIs by solving the Teukolsky
equation~\cite{Teukolsky:1973ha,Sasaki:1981sx,Finn:2000sy}. This
method is also used in \cite{Gair:2008bx} to estimate the EMRI rate
for LISA. Although the orbits of EMRIs generically have moderate
eccentricity and are inclined, the result consistently falls between
those from AKK and AKS, as shown in~\cite{Babak:2017tow}.

In this letter, we adopt the result from \cite{Finn:2000sy} where
the GW strain is organized into a set of harmonics $h_{c,m}(f)$ with
$m$ the harmonic number,
\begin{eqnarray}\label{eq:CharStrain}
&&h_{c,1} = \frac{5}{\sqrt{672\pi}} \frac{\eta^{1/2}M}{r_o} \tilde{\Omega}^{1/6} \mathcal{H}_{c,1}\ ,
  \nonumber \\
&&h_{c,m} = \sqrt{\frac{5(m+1)(m+2)(2m+1)! m^{2m}}{12\pi (m-1) [2^m
m!(2m+1)!!]^2}}
\frac{\eta^{1/2}M}{r_o}  \nonumber \\
&& \hspace{1.5cm} \times \tilde{\Omega}^{(2m-5)/6}
\mathcal{H}_{c,m}, \quad m\ge 2\ . \label{eq:hcm}
\end{eqnarray}
The equations are in geometrized units ($G=1$ and $c=1$). Here
$\eta$ is the ratio of the inspiraling object mass $\mu$ and SMBH
mass $M$, i.e. $\eta = \mu/M$. $r_o$ is the distance from the merger
to us. A dimensionless orbital angular velocity $\tilde{\Omega}$ is
defined as $\tilde{\Omega} \equiv M \Omega = 1/(\tilde{r}^{3/2}+s)$
where $\tilde{r} \equiv r/M$ with $r$ being the Boyer-Lindquist
radial coordinates of the orbit. $\mathcal{H}_{c,m}$ is the
relativistic correction and is provided in ~\cite{Finn:2000sy} with
various choices of $s$ and $r$ .

\begin{figure}
   \includegraphics[width=0.9\columnwidth]{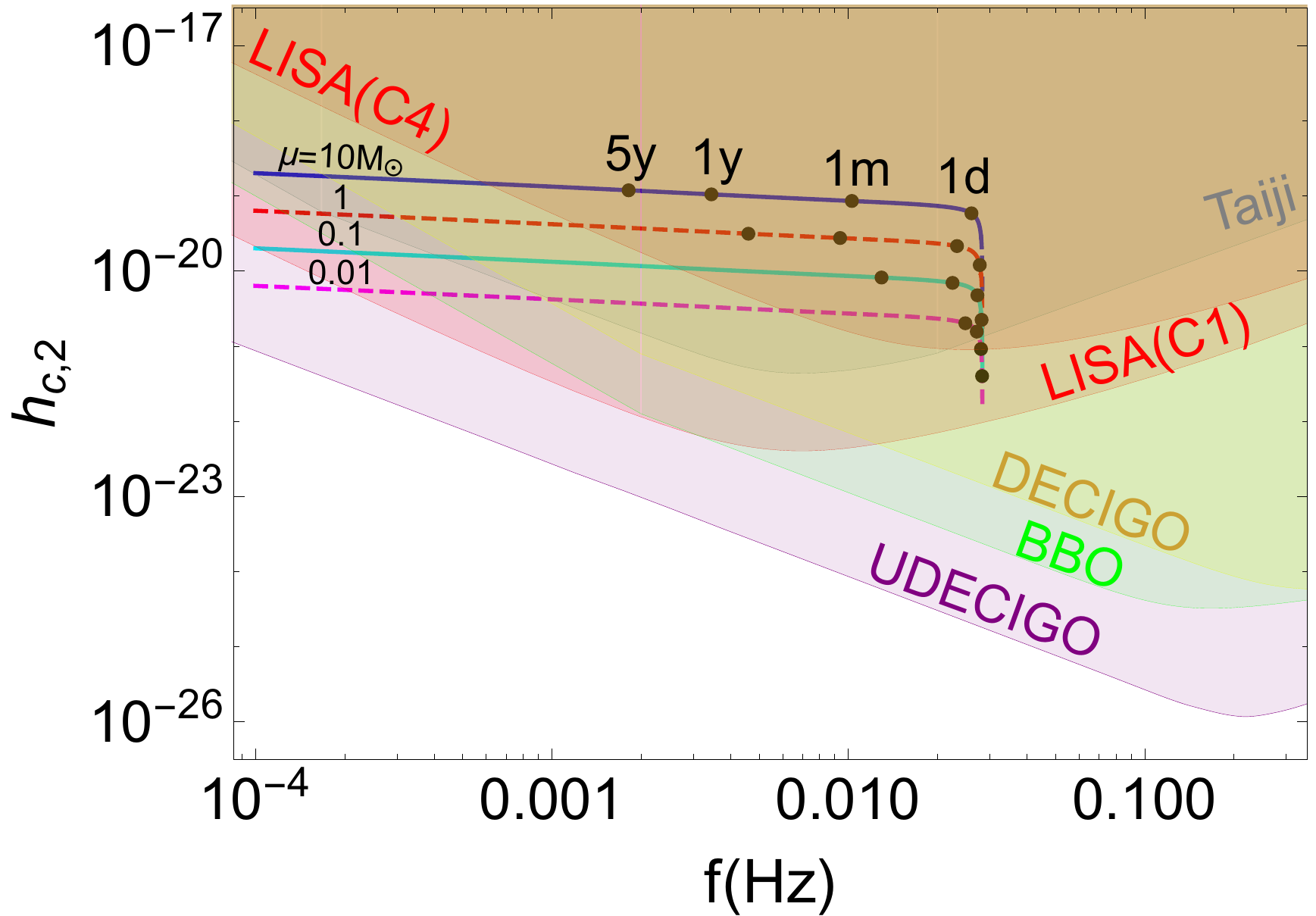}
  \caption{
\label{fig:vary-mBH} The characteristic strain $h_{c,2}$ is plotted
for different choices of PBH mass $\mu$. The SMBH has mass and spin
as $10^6 M_{\odot}$ and $0.999$. The distance to the earth is taken
to be $1\text{Gpc}$.  The dots indicate the remaining time before
the merger. The sensitivities of various proposed experiments,
$h_n(f)$, are also presented. }
\end{figure}

The maximal frequency of GW radiation $f_\text{max}$ occurs at the
innermost stable circular orbit (ISCO) at radius $r_{\text{ISCO}}$,
which is a function of $M$ and $a$~\cite{Bardeen:1972fi}. In
Fig.~\ref{fig:vary-mBH}, we show $h_{c,2}$ with different choices of
$\mu$. The experimental sensitivity is quantified by $h_n(f_m)
\equiv \sqrt{f S_n(f_m)}$, where $S_n(f_m)$ is the one-sided noise
power spectral density~\cite{Moore:2014lga}. Optimistic and
pessimistic LISA configurations N2A5M5L6 (C1) and N1A1M2L4
(C4)~\cite{Klein:2015hvg} are presented ~\footnote{We note that the
sensitivity curves LISA-C1 and LISA-C4, taken from
Ref.~\cite{Caprini:2015zlo}, are obtained from
LISACode~\cite{Petiteau:2008zz}. They are slightly different from
the LISA sensitivity presented in Ref.~\cite{Moore:2014lga}.}. We
also include several other proposed experiments, i.e. Taiji GW
project, Big Bang Observer (BBO), DECi-hertz Interferometer
Gravitational wave Observatory (DECIGO)~\cite{Moore:2014lga}, and
Ultimate-DECIGO (UDECIGO)~\cite{Kudoh:2005as}.

It is instructive to make a qualitative comparison between LISA and
LIGO at this point. While LISA and LIGO have their best
sensitivities at different frequency regimes, $h_n$ of LISA and LIGO
are at a similar order of magnitude. Around $r_{\textrm{ISCO}}$,
$h_c$ scales as $\sqrt{\mu M}$. The events observed by LIGO have
masses as $\mathcal{O}(10)$ $M_\odot$. At the same distance, a
similar order of magnitude of $h_c$ can be achieved if $\mu \sim
10^{-3} M_\odot$ when $M\sim 10^6 M_\odot$. This indicates the
possibility for LISA-like GW detectors to probe light PBHs.

A GW signal can be detected only if the SNR is above a certain
threshold. The SNR can be calculated as
\begin{eqnarray}\label{eq:SNR}
\text{SNR}^2 = \frac{\mathcal{S}^2}{\mathcal{N}^2} = \sum_{m} \int
\left[ \frac{h_{c,m}(f_m)}{h_n(f_m)} \right]^2 d \ln f_m,
\label{eq:snr}
\end{eqnarray}
where $\mathcal{S}$ and $\mathcal{N}$ are the signal and noise
obtained with matched-filtering~\cite{Moore:2014lga}. A widely
adopted choice of threshold is $\text{SNR} \ge 15$.

One subtlety appears when calculating SNR. While the slow inspirals
may last for a very long time, e.g. $\mathcal{O}(\textrm{Gyr})$,
LISA-like GW detectors can only operate at timescales
$\mathcal{O}(\textrm{yr})$. The GW frequency increases during
inspiral and achieves its maximal value $f_{max}$ when $r\sim
r_{\textrm{ISCO}}$, after which the inspiral stops and the plunge
occurs. Only a finite frequency window near the maximal frequency
can be recorded during the operation time of an experiment. A
truncation needs to be imposed accordingly for the integration range
in Eq.(\ref{eq:SNR}). This can be calculated by the total time
remaining before the plunge~\cite{Finn:2000sy,MTWgravitation}
\begin{equation}\label{eq:Time}
T=\frac{5}{256}\frac{1}{\mu}\frac{M^2}{{\tilde \Omega}^{8/3}} \mathcal
{T},
\end{equation}
where $\mathcal {T}$ is the general relativistic correction with
details listed in~\cite{Finn:2000sy}. Since we are focused on the
merger events, setting $T$ to the operation time gives the lower
bound of the frequency integral $f_{min}$. Note for smaller PBH
masses, the integration range can be very small since GW radiation
power is lower for a lighter CO. Thus light PBHs linger around ISCO
for a longer time and the frequency variation is tiny on timescales
$\mathcal{O}(\textrm{yr})$. For light PBHs, the variation of
frequency during $\mathcal {O}(\textrm{yr})$ is small. In this
limit, $\Delta f/ f\sim \mu/M^2$. Thus for a fixed $\mu$, a lower
$M$ provides a larger integral range when calculating SNR.

For each EMRI, the SNR imposes an upper limit on redshift. Combined
with the truncation imposed in the previous section, the limit of
the spatial integral is determined by $z_{\text{max}} =
min(z|_{\text{SNR}=15},1)$.

\vspace{0.2cm}

\noindent{\bfseries Intrinsic EMRI Rate for PBH-SMBH.} A CO can
change its orbit in two ways: $i)$ gravitationally scattering with
another CO object, or $ii)$ losing energy by GW radiation. If
gravitational scattering brings a CO to an orbit direct falling into
a SMBH, this plunge wil not produce a GW observable by LISA-like
detectors. On the other hand, if a SMBH-CO merger is induced by GW
radiation after many orbits, this results in a slow inspiral which
can be potentially detected. This will be our focus. \footnote{Using
LISA to detect the GW radiation from PBH inspiraling into
Sagittarius A* is also discussed in \cite{Kuhnel:2017bvu}. This
paper considers the GW emission before merging, which is the
stochastic signal from the extreme long time ongoing inspirals.
However, as we will discuss later, the scatterings between PBH and
other stellar objects can easily change PBH's orbit, which is
crucial for the events counting. Neglecting them will largely
overestimate the event rate. Further the GW frequency distribution
spans a large range and needs to be studied more carefully in order
to properly calculate SNR.}

The intrinsic EMRI rate induced by SMBH-stellar BH slow inspiral has
been calculated using the Fokker-Planck equation
in~\cite{Hopman:2006qr,Hopman:2006xn,Hopman:2009gd}. The stellar BH
mass is set to be $10 M_\odot$, and the number density is taken to
be $0.1\%$ of the total number density of astrophysical objects
within $r_h$. It can be explicitly written as~\cite{Hopman:2009gd}
\begin{equation}\label{eq:StellarBHNumDen}
  n_{\text{BH}} = 40\ \text{pc}^{-3} \left( \frac{M}{3 \times 10^6 M_{\odot}} \right)^{-1/2}.
\end{equation}
The intrinsic EMRI rate of such system scales with $M$
as~\cite{Hopman:2009gd,Gair:2008bx}
\begin{eqnarray}\label{eq:StellarBH}
  \mathcal{R}_{\text{astro}}(M) = 400\text{Gyr}^{-1} \left(\frac{M}{3\times
10^6 M_{\odot}}\right)^{-0.15} . \label{eq:rate}
\end{eqnarray}
Now we study how Eq.(\ref{eq:StellarBH}) scales as a function of PBH
number density and mass.

First, we rescale the number density of PBHs with respect to that of
stellar BHs in Eq.~(\ref{eq:StellarBHNumDen}),
\begin{eqnarray}
  \mathcal{G}(M,\mu) = f_{\textrm{PBH}}\frac{\rho_{\text{NFW}}(M,r_h(M))/\mu}{n_{\text{BH}}(M)} .
\end{eqnarray}
For example, when $\mu=10 M_{\odot}$ and $M= 10^6 M_{\odot}$,
$\mathcal{G}$ is $\mathcal {O}(1)$.

The timescale that brings a PBH to an orbit of slow inspiral can be
written as a function of relaxation time $t_h$ at $r_h$. According
to~\cite{MiraldaEscude:2000cg}, for generic astrophysical objects,
the relaxation time is determined by the species with largest $m_i^2
n_i$ where $m_i$ and $n_i$ are the mass and number density of each
species. Using the NFW profile, the total mass of the PBH within
$r_h$ is only a small fraction. Given the parameter choice in
\cite{Hopman:2009gd}, the relaxation of PBHs is mainly controlled by
their scattering with main-sequence stars (MS). Accordingly we
expect $t_h$ is approximately independent of PBH mass.

The angular momentum relaxation time can be written as
\begin{eqnarray}
  t_J(J, a) = t_h \left[\frac{J}{J_m(a)}\right]^2 \left(\frac{a}{r_h}\right)^p\ .
\end{eqnarray}
Here $a$ is the semi-major axis of an orbit, and $J_m(a)=\sqrt{M a}$
is the maximal (circular) angular momentum for a specific energy.
$p$ is related to the spatial profile of the astrophysical objects
which dominate the relaxation process of PBHs, i.e.
$n_{\textrm{MS}}\sim r^{-3/2-p}$.

Now let us estimate the timescale of a slow inspiral. This process
lasts a long time, much longer than the period of the orbit. The
energy carried away by gravitational radiation per period
is~\cite{Hopman:2005vr,Peters:1963ux}:
\begin{eqnarray}
\Delta E  = E_1 \left(\frac{J}{J_{lc}}\right)^{-7}
\end{eqnarray}
with
\begin{eqnarray}
E_1=\frac{85\pi}{3\times 2^{13}} \frac{\mu}{M}; \ \ \ J_{lc}=4 M.
\end{eqnarray}
Note the energy and angular momentum are defined in units of PBH
mass $\mu$.

For an orbit with high eccentricity, periapse approximately remains
a constant, and the time for a CO with initial specific energy
$\epsilon_0$ to finish the inspiral is
\begin{eqnarray}
  t_0 = \int_{\epsilon_0}^{\infty} \frac{d \epsilon}{d\epsilon/dt}
  \approx \frac{2\pi\sqrt{ M a}}{\Delta E}\sim \mu^{-1} .
\end{eqnarray}
Here we only pay attention to its dependence on $\mu$ since the goal
is to estimate the intrinsic EMRI rate by rescaling
Eq.(\ref{eq:StellarBH})

It is important to ensure that the slow inspiral can continue
without being disrupted by further scatterings. A critical value of
$a$ is defined by the ratio of $t_0$ and $t_J$, i.e.
$t_0(J_{lc},a_c)/t_J(J_{lc}, a_c)=1$. For an orbit with $a<a_c$, a
CO has a large chance to fall into SMBH without disruptions. This
critical value $a_c$ is given by,
\begin{eqnarray}
  \frac{a_c}{r_h} =\left(\frac{d_c}{r_h}\right)^{\frac{3}{3-2p}} ;\ \ \  d_c =
\left(\frac{8 \sqrt{M} E_1 t_h}{\pi}\right)^{2/3} .
\end{eqnarray}
Using the analytic solution of the Fokker-Planck equation
in~\cite{Hopman:2005vr}, one obtains an estimation of the intrinsic
EMRI rate for PBHs with arbitrary mass,
\begin{eqnarray}
\label{eq:p-scaling}
\mathcal{R}_{\text{PBH}}(M,\mu)
  &=& \int_0^{a_c} \frac{d a\ n_{\text{PBH}}(a)}{\ln(J_m(a_c)/J_{\text{lc}})t_h}\left(\frac{r_h}{a}\right)^p \nonumber \\
  &\sim& \frac{n_{\textrm{PBH}}(r_h)}{t_h \textrm{ln}[J_m(a_c)/J_{lc}]}\left(\frac{a_c}{r_h}\right)^{3/2-2p}\nonumber\\
                           &\sim& \mathcal{G}(M,\mu)\ \mu^{\frac{4p-3}{2p-3}}\ \mathcal{R}_{\text{astro}}(M).
\end{eqnarray}
where $n_{\text{PBH}}(a)$ is the PBH number density at $a$
\footnote{For simplicity, we assume PBH and MS share the same power
law, i.e. $p$, as spatial distribution. This is reasonable when PBH
is lighter or comparable to $\mathcal {O}(1)\ M_\odot$. It is not
difficult to derive a similar formula with different choices on
$p$.}.

As shown in Eq.(\ref{eq:p-scaling}), the intrinsic EMRI rate is
sensitive to the choice of $p$, which ranges from 0 to 0.25
~\cite{Hopman:2006xn,AmaroSeoane:2010bq,Alexander:2008tq,Preto:2009kd}.
To show its effects qualitatively, we present the results with
different choices of $p$ in the next section.

\vspace{0.2cm}

\noindent{\bfseries PBH Constraints.} Finally, to estimate event
rate, we take the mass spectrum of SMBHs given in
Ref.~\cite{Klein:2015hvg,Babak:2017tow},
\begin{eqnarray}
\label{eq:dndM}
  \frac{dn}{d\ln M} = 0.005 \left(\frac{M}{3\times
10^6M_{\odot}}\right)^{-0.3} \text{Mpc}^{-3},
\end{eqnarray}
with the range of the SMBH masses taken to be $10^4 M_{\odot} \le M
\le 10^7 M_{\odot}$. One can convert the expected observable
PBH-SMBH EMRI rate into the sensitivity to PBH energy density
fraction of DM, $f_\text{PBH}$.

Once such EMRI events are observed, the detailed waveform provides
an excellent handle to extract information on the
system~\cite{Barack:2003fp,Babak:2017tow}, and $\mu$ can be measured
by analyzing the time-dependence of the orbit. The stellar BHs are
expected to have masses ranging from $5$ to few tens
$M_{\odot}$~\cite{Belczynski:2009xy}. If PBHs are within the same
mass regime, e.g. motivated in~\cite{Georg:2017mqk}, stellar BHs may
behave as a background of the PBH search. Further, mergers between
SMBH and other astrophysical COs, such as neutron stars and white
dwarfs, may also contribute as PBH-SMBH background. The mass of
white dwarfs (neutron stars) is unlikely to be smaller than $0.6\
M_\odot$ ($1\ M_\odot$). If PBHs are much lighter than those
astrophysical COs, the background is free. In that case, one event
observed is enough to declare discovery.

\begin{figure}
\includegraphics[width=0.99\columnwidth]{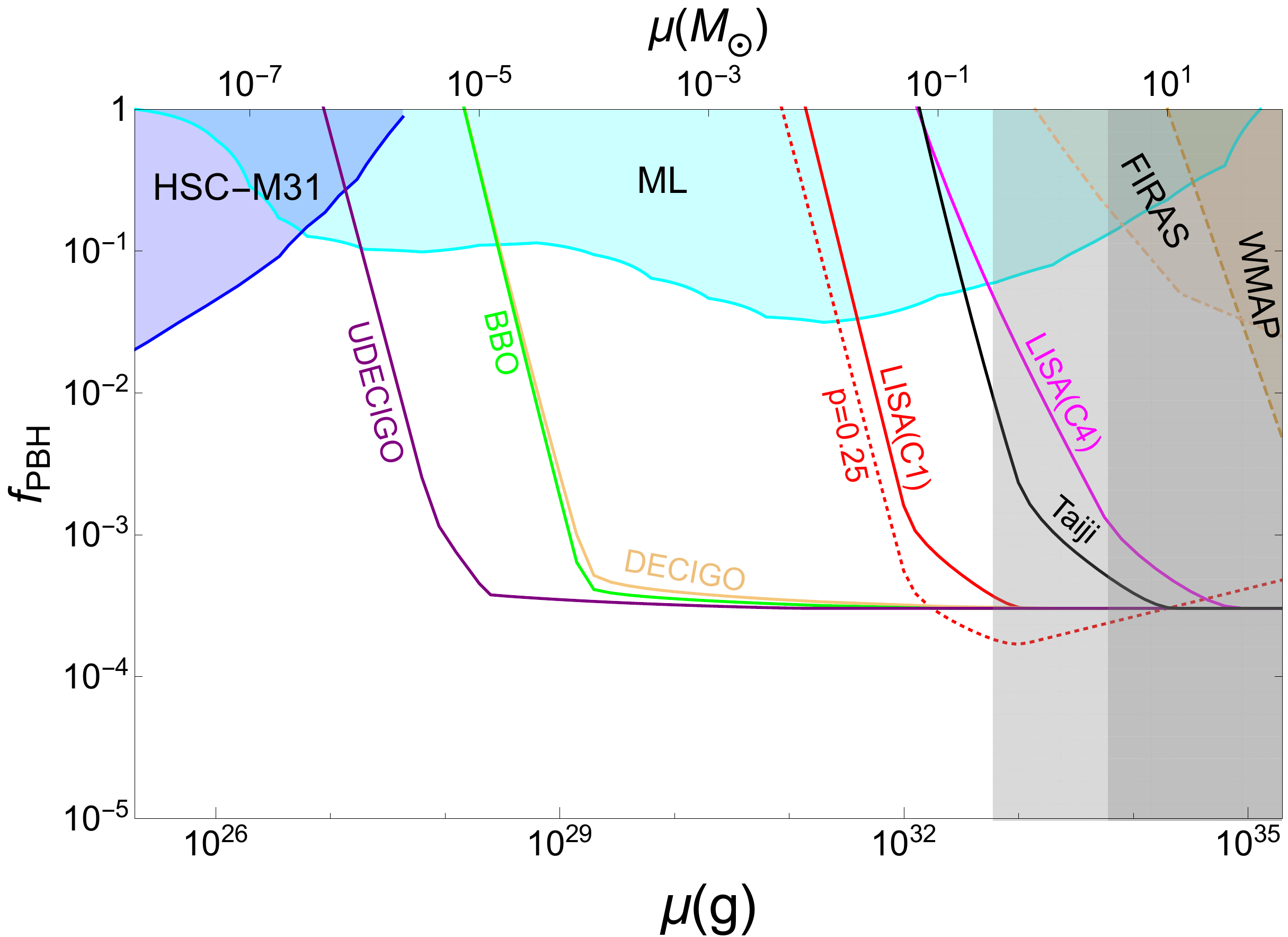}
\caption{We show the value of $f_{\textrm{PBH}}(\mu)$ which is
expected to give one observable PBH-SMBH EMRI event during the
5-year mission of an experiment. Various detector configurations and
sensitivities are considered.  The solid lines are obtained by
taking $p=0$, and the dashed red line corresponds to the LISA C1
sensitivity with $p=0.25$. The microlensing constraint, HSC-M31, is
from Ref.~\cite{Niikura:2017zjd}, and other constraints are from
Ref.~\cite{Carr:2016drx}. The regions where $0.3 M_{\odot} < \mu < 3
M_{\odot}$ and $3 M_{\odot} < \mu < 100 M_{\odot}$ are shaded. Here
the background from neutron star (white
dwarf)~\cite{Kiziltan:2013oja,whitedwarfmass} and stellar BHs,
respectively, needs to be carefully considered. \label{fig:PBH} }
\end{figure}

In Fig.~\ref{fig:PBH}, with various choices of GW detectors, we
present the value of $f_{\text{PBH}}$ which generate one PBH-SMBH
EMRI with SNR $> 15$ during a 5-year operation of the experiment.
The dark grey region starts at 3 $M_\odot$ where stellar BHs begin
to contribute as background. From 0.3 $M_\odot$, white dwarfs and
neutron stars become important. We stop our calculation at
$\mu=10^{2} M_\odot$ so that EMRI remains a reasonable
approximation, especially for galaxies with light SMBHs ($10^4
M_\odot$). The existing constraints on $f_\text{PBH}$ are included,
and LISA-like GW experiments have good potential to probe the
unexplored parameter space.

There are several important features of this sensitivity curve.

i). When $\mu$ is not too small, with a sufficiently sensitive GW
detector, all EMRIs happening within $z=1$ can be observed. As
indicated in Eq. (\ref{eq:p-scaling}), the intrinsic EMRI rate
$\mathcal{R}_\text{PBH}(M, \mu)$ is independent of $\mu$ when $p=0$.
This explains the flatness of $f_{\textrm{PBH}}$ curves in the large
$\mu$ regime. When lowering $\mu$, not all EMRIs exceed the SNR
threshold. This produces the turning point which is determined by
the detector sensitivity.

ii). As discussed below Eq.(\ref{eq:Time}), for a fixed $\mu$,
smaller $M$ gives a larger integration range of $\Delta f/f$ in the
calculation of SNR, i.e. $\Delta f/f\sim 1/M^2$. Although the
gravitational wave strain scales as $h_c\sim \sqrt{M}$, a better SNR
can still be achieved for lighter SMBH assuming $h_n$ is the same.
Given the SMBH mass distribution also increases when $M$ decreases
as shown in Eq.(\ref{eq:dndM}), this indicates that a GW experiment
may have better sensitivity for lighter PBHs if its best frequency
region is higher. This is why the reach of DECIGO is comparable to
that of BBO even though its sensitivity is worse in lower frequency.

In Fig.~\ref{fig:PBH}, we also study the reach limit with a
different choice of $p$, shown as the dashed curve for LISA(C1). For
$p\neq0$, the dependence on $\mu$ becomes non-trivial for the
intrinsic EMRI rate. When $p$ is positive, the probed region is
further extended in the lighter PBH region. As discussed above, $p$
is related to the spatial distribution of the astrophysical objects,
presumably MS, and controls the relaxation time. It also affects the
EMRI rate of merging SMBHs and ordinary astrophysical COs, the
observation of which can help to reduce the uncertainty in our
PBH-SMBH rate calculation.


\vspace{0.2cm}

\noindent{\bfseries Discussion.} In this letter, we explore the
possibility of using LISA-like GW detectors to look for PBH-SMBH
EMRI events. The frequency of the GWs is mainly determined by the
mass of SMBH, and a vast range of PBH masses can be probed by such
experiments. Especially, a BH much lighter than 0.3 $M_\odot$ is not
expected from astrophysics. The detection of such a SMBH-PBH merger
outside the astrophysical CO mass window is potentially enough to
declare the discovery of PBHs.

We find that LISA-like GW experiments provide a novel and promising
way to test the scenario where PBHs are (a fraction of) DM. The
sensitivity to $f_\text{PBH}$ in certain mass regimes could be as
good as $10^{-3} \sim 10^{-4}$, which is much better than the
existing constraints.

Our analysis here initiates the study of PBHs as DM using LISA-like
GW detectors which connects astronomy and GW and DM physics. We
expect that our current results can be significantly improved with
better knowledge from those interdisciplinary areas in the future.
For example, we truncate our calculation at $z=1$ due to the
uncertain validity of astrophysical empirical relations, such as
Eq.(\ref{eq:DMProfile}) at high redshift. With a better
understanding of such a relation, the higher $z$ region could be
included, and a much smaller $f_\text{PBH}$ may be explored.
Furthermore, astrophysical uncertainties, such as mass and spin
distributions of SMBHs, would affect the rate estimation. The
observation of EMRI events induced by astrophysical COs also
provides valuable information. This may have feedback to the PBH
calculation and reduce the theoretical uncertainties.

As a final comment, as we discussed above, lighter SMBHs may
potentially be more beneficial to search for small mass PBHs, both
because of the higher number density from the SMBH mass spectrum as
well as the larger integration window on frequency in the SNR
calculation. This serves as a guideline for the optimization of a
light PBH search in future LISA-like GW experiments.



\noindent{\bfseries Acknowledgement.} We would like to thank Yanbei
Chen, Runqiu Liu, Aaron Pierce, Tao Ren,  Keith Riles and Ben Safdi
for helpful discussions. Especially, we thank Xian Chen, Josh Foster
for carefully reading our draft and giving valuable comments. JS is
supported by the National Natural Science Foundation of China (NSFC)
under grant No.11647601, No.11690022 and No.11675243 and also
supported by the Strategic Priority Research Program of the Chinese
Academy of Sciences under grant No.XDB21010200 and No.XDB23030100.
YZ thank the support of grant from the Office of Science and
Technology, Shanghai Municipal Government (No. 16DZ2260200). YZ is
also supported by US Department of Energy under grant DE-SC0007859.


\end{document}